\documentstyle[aps,12pt]{revtex}

\begin{document}

\title{Magnetic susceptibility, specific heat and dielectric
constant of hexagonal YMnO$_3$, LuMnO$_3$ and ScMnO$_3$}

\author{D.G. Tomuta, S. Ramakrishnan\dag , G.J. Nieuwenhuys and J.A. Mydosh}

\address{Kamerlingh Onnes Laboratory, Leiden University, 2300 RA Leiden, The Netherlands}
\address{ and \dag\ Tata Institute of Fundamental Research, Mumbai-400005, India }
\maketitle
\begin{abstract}

We report the magnetic susceptibility, specific heat and
dielectric constant on high purity polycrystalline samples of
three hexagonal manganites: YMnO$_3$, LuMnO$_3$ and ScMnO$_3$.
These materials can exhibit a ferroelectric transition at very
high temperatures (T$_{FE}$ $>$ 700~K). At lower temperatures
there is magnetic ordering of the frustrated Mn$^{3+}$ spins (S=2)
on a triangular Mn lattice (YMnO$_3$~:~T$_N$=~71~K;
LuMnO$_3$~:~T$_N$=~90~K and ScMnO$_3$~:~T$_N$=~130~K). The
transition is characterized by a sharp kink in the magnetic
susceptibility at T$_N$ below which it continues to increase due
to the frustration on the triangular lattice. The specific heat
shows \emph{one} clear continuous phase transition at T$_N$, which
is independent of external magnetic field up to 9~T with an
entropy content as expected for Mn$^{3+}$ ions. The temperature
dependent dielectric constant displays a distinct anomaly at
T$_N$.
\end{abstract}


\newpage

\section{Introduction}
Rare-earth manganites of the perovskite-type structure REMnO$_3$
have been discovered in the 1950s. Materials with small ionic
radius (RE = Ho, Er, Tm, Yb, Lu, Y and Sc) crystallize in the
hexagonal structure [S.G.\emph{P6$_3$cm}] while the compounds with
larger ionic radius (RE = La, Ce, Pr, Nd, Sm, Eu, Gd, Tb or Dy)
are orthorombic [S.G.\emph{Pnma}]~\cite{Yakel}. The orthorhombic
compounds (S.G. \emph{Pnma}) form the basis for the CMR materials
and they have been studied widely in the recent years. Unlike
their orthorhombic counterparts, investigations in hexagonal
REMnO$_3$ compounds are few despite their interesting properties.
Systems belonging to the hexagonal class undergo a ferroelectric
transition far above the ordering of Mn$^{3+}$ spins. Such
magnetic ordering occurs in both structures, however,
ferroelectricity is possible only for the hexagonal phase which
has the noncentrosymmetric phase group \emph{P6$_3$cm}. Although
hexagonal manganates have been studied for many
years~\cite{koehler,WA,McCarty,smolenski}, very recently there has
been a resurgence of interest in these
materials~\cite{fujimura,Bieringer99,Munoz00}. Most hexagonal
REMnO$_3$ compounds can exhibit two transitions, i.e., a very high
temperature (T$_{FE}$~$\approx$ 900~K) ferroelectric distortion
and a low temperature ($\approx$ 100~K) magnetic ordering. The
development of two order parameters is a rarity in
oxides~\cite{Hill00} and opens the possibility for
electric-magnetic interactions and "tuning" with electric and
magnetic fields.

Current efforts have focused on the magnetic transition and
refining the magnetic structure~\cite{Bieringer99,Munoz00,wan}.
These investigations have been complicated by the formation of a
ferromagnetic impurity phase, Mn$_3$O$_4$, with a Curie
temperature T$_C$ $\simeq$ 43~K. Amounts of order of 0.5 at.\% of
this phase are found in most samples reported to
date~\cite{koehler,Munoz00,Xu}, causing significant anomalies in
the bulk properties. For example, the true REMnO$_3$
antiferromagnetic transition is masked by the ferromagnetic signal
in the susceptibility and the specific heat may exhibit a second
strongly magnetic field dependent peak around 40~K with a tiny
entropy content.

In order to examine the intrinsic magnetic transition, we have
developed a preparation technique for fabrication of pure samples
(impurity content less then 0.1 at.\%) of YMnO$_3$,  LuMnO$_3$ and
ScMnO$_3$. We have measured the magnetic susceptibility~($\chi$)
up to 400~K in different external fields. Specific heat~(C$_p$)
and its field dependence were also determined. Using a two
Debye~-~temperature model we can deduce the
excess~$\triangle$C$_p$ and the magnetic entropy. Finally, we have
measured the dielectric constant~($\epsilon$). Our results show
that there is a single antiferromagnetic transition at T$_N$ in
each of these compounds. However, there appears a continuous
increase in $\chi$(T) and a broad low T maximum ($\approx$ 50~K)
in $\triangle$C$_p$/T with significant entropy as the temperature
is reduced. We attribute these anomalous features at T~$<$~T$_N$
to the frustration of Mn$^{3+}$ spins on a triangular lattice.
From our measurements of the dielectric response of our high
purity samples we find a more enhanced "S"-shaped curve than the
one reported in a previous investigation~\cite{Huang}. This
signature of T$_N$ in the dielectric susceptibility at T$_N$ $\ll$
T$_{FE}$ we believe is caused by a small change in the
ferroelectric domain wall mobility and {\bf not by a direct
coupling} between the magnetic and ferroelectric order parameter.

\section{Sample preparation and Experimental techniques}
\label{two}
 The samples investigated were prepared by the
solid-state reaction technique at ambient pressure. Cation oxides
of Y$_2$O$_3$, Lu$_2$O$_3$, Sc$_2$O$_3$ (99.99\%) and MnO$_2$
(99.99\%), obtained from Alpha Aesar~\cite{AA}, were mixed in a
1:2 molar ratio to achieve the stochiometry of hexagonal
LnMnO$_3$. The mixture was well ground and calcinated in
O$_2$-flow at 1100$^0$C for 24h. To ensure better homogeneity  the
mixtures were ground again, compacted into small pellets and
reheated in air at 1400$^0$C for 48 h for YMnO$_3$ and LuMnO$_3$;
and at 900$^0$C in flowing O$_2$ for ScMnO$_3$. It is only when
this high-temperature "reheating" is done for YMnO$_3$ and
LuMnO$_3$, and the exact procedures given in ref.~\cite{karen98}
are followed for ScMnO$_3$ that we obtain impurity-free samples.
This optimum procedure unfortunately depends slightly on the
starting materials but the magnetic and structural properties of
these compounds seem not to be sensitive to the O$_2$ content. We
may have a small $\delta$ missing/excees in the oxygen
stochiometry such as REMnO${_{3\pm\delta}}$.

The powder x-ray diffraction data on all three compounds show that
the polycrystalline samples have the correct hexagonal structure
without any trace of impurity phases. The lattice parameters found
for each compound are in agreement with previous
reports~\cite{koehler,McCarty,fujimura}. The radius of the pellets
is 1.5 mm and their thickness varies from 0.5~mm to
 0.8~mm.

Dc-susceptibility ($\chi$) was measured in a Quantum Design
magnetometer (MPMS~5S) for each sample using two fields, 0.1~T and
2~T in the temperature range from 1.8 to 400~K. Heat-capacity
measurements were performed using a Quantum Design PPMS system in
the range from 1.8 to 250~K and in fields up to 9~T. A relaxation
technique was used with a resolution of 2\% and an accuracy better
than 5\%. All samples showed only \emph{one} sharp anomaly in the
specific heat at T$_N$ without any additional peaks that could be
ascribed to an impurity phase. Comparing with the specific heat of
the most common impurity (Mn$_3$O$_4$) this leads to an estimation
of less than 0.05\% of this impurity. We also estimated the amount
of second phase in all samples by measuring magnetization in
different fields, which leads to even lower amounts. Finally we
measured the dielectric constant $\epsilon$ as function of the
temperature using a capacitance bridge operating at a frequency of
1 MHz.

\section{Susceptibility}
Magnetic susceptibility measurements on the hexagonal compounds
previously investigated failed to show any clear anomaly at T$_N$
in the $\chi$ versus $T$ curve. In \ref{fig.TOMU01},
\ref{fig.TOMU02} and  \ref{fig.TOMU03} we show the results on our
high purity samples. The anomaly at T$_N$ is evident from the
graphs. Reciprocal susceptibilities are also shown as an inset of
these figures.

As expected for triangular antiferromagnets the susceptibility at
the lowest temperature (T$<$ T$_N$) does not decrease to 2/3 of
its value at T$_N$ as for the classical two-sublattice case. From
the earliest neutron diffraction combined with group theory
arguments for this hexagonal lattice, Berthaut {\it et
al.}~\cite{alpha} concluded that the Mn magnetic moments are
directed in the basal plane and are oriented with angles of
120$^o$ with respect to each other. If the field is perpendicular
to the magnetic plane all spins will be canted out from the plane
with a small angle $\delta$ in the direction of the applied field.
For this field direction the zero-temperature susceptibility will
be equal to $\chi(T_N)$. A simple calculation with classical spins
arranged with 120$^o$ angles, shows that $\chi(T=0)$ is at least
$\chi(T_N)$ for all directions of the field in the basal plane. A
detailed calculation will be carried out for this intriguing
behavior.

The insets of the figures \ref{fig.TOMU01}-~\ref{fig.TOMU03}
exhibit the reciprocal susceptibility versus temperature. By
fitting with a Curie-Weiss law in the temperature range from 300
to 400~K the values of the effective Mn moments ($\mu_{eff}$) and
the paramagnetic Curie temperatures ($\theta_{CW}$) could be
obtained. \ref{tab1} collects these parameters together with
T$_N$, taken as the temperature where the kink in $\chi$ versus T
occurs.

Because of the large values of $\theta_{CW}$ the temperature range
for the fits (300~K - 400~K) is still too low. This explains the
slight discrepancies between the obtained effective moments and
the expected value of the $Mn^{3+}$(4.9$\mu_B$). Also the absolute
values of $\theta_{CW}$ are much larger then the ordering
temperature $T_N$. Ramirez~\cite{ramirez} defined the ratio
between these parameters as the measure of the geometrical
frustration of the antiferromagnetic system. For our in-plane
triangular lattice these value appears to be of order 5 which puts
our hexagonal system  in the class of moderately frustrated
systems.

Note that a ferromagnetic impurity (Mn$_3$O$_4$) content of less
then 0.05\% will be immediately visible from the susceptibility
measurements in two fields (not shown), e.g. 0.1 and 2~T. Our
samples do not show any sign of the presence of Mn$_3$O$_4$
meaning that its abundance is less then 100 ppm.

\section{Specific heat}
The results of the specific heat measurements for all three
compounds in 0 and 9 T magnetic fields are shown in the insets of
the \ref{fig.TOMU04}, ~\ref{fig.TOMU05} and \ref{fig.TOMU06}
together with the estimates for their lattice contribution (dash
lines). The lattice specific heat was estimated using two Debye
functions for the "heavy"(Mn+RE) and "light"(O$_2$) elements in
our compounds. The fits were carried out considering the specific
heat at temperatures above 1.5 times $T_N$. We found reasonable
values for the Debye temperatures, i.e.,
$\theta_D$("light")$\approx$ 780~K - 800~K and $\theta_D$("heavy")
$\approx$ 350~K - 420~K. Subtraction of this estimation of the
lattice specific heat provides us the (excess) magnetic
contribution of the system. These $\Delta$C$_p$ are shown as
$\Delta$C$_p$/T versus T curves in
\ref{fig.TOMU04},~\ref{fig.TOMU05} and \ref{fig.TOMU06} (left
y-axis). On the right ordinate we plot the entropy computed from
the integral. All samples exhibit sharp anomalies at the
corresponding transition temperature indicating the onset of
antiferromagnetic order. Up to an applied field of 9~T the
specific heat is independent of the field. Contrary to previous
work~\cite{Munoz00} we could not detect a second lower temperature
anomaly, which becomes visible when the sample is contaminated by
more then 0.5\% of impurity (Mn$_3$O$_4$). The resulting entropy
is in excellent agreement with that expected for a Mn$^{3+}$ ion
(S=2).
\begin{displaymath}
\Delta S = R\cdot {ln(2S+1)}= 13.38~J/mol\cdot K
\end{displaymath}

\noindent In the literature~\cite{Munoz00,Xu} discrepancies
between the deduced and expected entropy up to a factor of 10 have
been reported due to an inaccurate estimate of the lattice
contribution. Our analysis gives values for $\Delta$S of at least
90\% of the expected one. Note that the magnitude of $\Delta$C$_p$
at $T_N$ is approximately the same for all three compounds
(18~J/mol~K).

Furthermore, there is an the interesting anomaly ("bump") at low T
($\approx 50~K$) in the excess specific heat ( $\Delta$C$_p$/T).
The amount of entropy removed directly at $T_N$ is less than half
of the total and this relative amount decreases going from Y via
Lu to Sc. The physical cause of this "bump" in $\Delta$C$_p$/T and
the corresponding entropy involved is related, in our opinion, to
the frustrated magnetism. Here it seems that there are still spin
degrees of freedom available far below T$_N$ and the spin system
is progressively searching for its true ground state. We believe
the "bump" is correlated with the constantly increasing
susceptibility below T$_N$ (see
figures~\ref{fig.TOMU04}~-~\ref{fig.TOMU06}).

 Our observations are in agreement with different
magnetic structures found via neutron diffraction by Mu\~{n}oz
$\textit {et~ al.}$~\cite{Munoz00}, via Raman-scattering and
infrared-active spectra studied by Iliev $\textit {et~
al.}$~\cite{iliev} and via non-linear optical spectroscopy by
Fr\"{o}hlich $\textit {et~ al.}$~\cite{frolich,frolichPRL}.
Thermal expansion measurements on single crystals are in progress
for these hexagonal frustrated systems in order to give us a
better picture of this "hidden" processes that occur at lower
temperatures.

\section{Dielectric susceptibility}

In general, one expects a coupling between ferroelectricity and
magnetism only when T$_{FE}$ is close to T$_N$. However, a
coupling between the ferroelectric and antiferromagnetic order
parameters in YMnO$_3$ has been recently claimed by Huang $\textit
et~ al.$~\cite{Huang} based on the observation of a small
"S"-shaped anomaly ($\sim$2\%) at the Neel temperature in the
$\epsilon$ versus $T$ curve. They have suggested that such a
coupling could arise due to the magnetostrictive effect associated
with a lattice change accompanying the antiferromagnetic ordering
of Mn$^{3+}$ spins. We also have measured the dielectric response
of our high~purity samples and our data reveal a more enhanced
"S"-shaped curve - see \ref{fig.TOMU07}. Note that the dielectric
susceptibility, $\epsilon $(T) is probed at T$_N$ $\ll$ T$_{FE}$
and that the step at T$_N$ is rather small. These leads us to
believe that we are observing a magnetic ordering effect on the
mobility of the domain walls of the ferroelectric domains and not
a direct coupling between the two order parameters.

\section{Discussions}

For samples containing small amounts of impurity phase
(Mn$_3$O$_4$ with T$_C$ = 43~K) strong anomalies occurs around
43~K. For example, sharp spikes with small entropy contents appear
in the specific heat~\cite{Munoz00} and there are strong "upturns"
and field dependence in susceptibilities. Our specially prepared
samples (see \ref{two}), whose data are shown in the previous
figures, are completely free of such anomalies. Therefore, we
believe our samples are without the troublesome ferromagnetic
impurity phase and the claims of unusual behavior around 40~K by
other authors are the result of impure samples.

There is remarkable difference in the bulk properties of YMnO$_3$
and LuMnO$_3$ on one hand and those of ScMnO$_3$ at the other. The
magnetic susceptibility of the former two compounds shows the
characteristic kink at T$_N$, while the susceptibility of
ScMnO$_3$ increases much more rapidly as temperature is decreased
through the Neel point- see \ref{fig.TOMU08}. Also, the magnetic
susceptibility of ScMnO$_3$ is sensitive to magnetic history of
the sample, i.e., there is a difference between the field cooled
and zero-field cooled values even for magnetic fields as low as
50~mT. Such field dependent history is not found in low field
$\chi$ measurement for the YMnO$_3$ nor for LuMnO$_3$.

Considering the specific heat at the lowest temperatures, one sees
that the total specific heat of ScMnO$_3$ is proportional to T$^2$
up to a temperature of 30~K, while those of YMnO$_3$ and LuMnO$_3$
show a much stronger T dependence (see inset of~\ref{fig.TOMU08}-
logarithmic scale). We would expect that the low temperature
specific heat of an antiferromagnetic material can be described by
a T$^3$ term due to phonons and antiferromagnetic spin-waves
(possibly with an exponential term due to a gap at zero
wave-vector in the spin-wave spectrum). Now, a T$^2$ dependence at
low temperatures (below 30~K) shows that the specific heat is
dominated by magnetic effects rather then by phonon excitations.
Such a magnetic contribution which deviates from the
antiferromagnetic spin-wave type indicates a ferromagnetic origin
where the specific heat due to spin waves should follow T$^{3/2}$.

These subtle differences in the magnetic behavior are the subject
of our neutron diffraction and muon spin rotation experiments on
polycrystalline and single crystal materials of these fascinating
Mn-oxides.
\section{Conclusions}

We have synthesized for the first time polycrystalline samples of
the hexagonal REMnO$_{3}$ compounds with a sufficient purity such
that the antiferromagnetic magnetic ordering is revealed in the
magnetic susceptibility. We have measured the specific heat and we
show that an adequate analysis leads to the full entropy of the
magnetic Mn-ions. We find that a considerable part of this entropy
is released only at temperatures of order of half of the Neel
temperature. There appears to be a remarkable difference between
the properties of YMnO$_{3}$ and LuMnO$_{3}$ on the one hand and
those of ScMnO$_{3}$ on the other. Finally, we have measured the
dielectric susceptibilities, which shows a kink at the respective
magnetic transition temperatures. We believe this to be due to an
influence from the antiferromagnetic ordering on the mobility of
the ferroelectric domain-walls.

  The authors gratefully acknowledge R.W.A. Hendrikx, and T.J.
Gortenmulder  for their technical assistance in analyzing the
samples, H.B. Brom and I.G. Romijn for the use of their dielectric
susceptibility apparatus and  R. Feyerherm for valuable
discussions. This work was supported by the Dutch Foundation FOM.

\newpage

\begin{table}[tbp]
\caption{\label{tab1}Effective Mn moments, Neel temperatures,
paramagnetic Curie-Weiss temperatures and frustration parameters.}
\begin{tabular}{@{}lllll}
  & $\mu_{eff} / \mu _{B}$ & $T_N$(K) & $\theta_{CW}$(K)
& $|\theta_{CW}|/T_N$

\\
 YMnO$_{3}$       & 4.91 & 71 &-417 & 5.9       \\

LuMnO$_{3}$     &4.78  & 90& -519 & 5.8    \\

ScMnO$_{3}$    &4.11& 130& -495& 3.8      \\
\end{tabular}
\end{table}

\newpage

\begin{figure}
\caption{\label{fig.TOMU01}Low
temperature susceptibility of YMnO$_3$ in a field of 2~T. The kink
represents the Neel temperature. Inset: inverse susceptibility and
the high-T Curie-Weiss fit (solid line).}
\end{figure}

\begin{figure}
\caption{\label{fig.TOMU02}Low temperature susceptibility of
LuMnO$_3$ in a field of 2~T. The kink represents the Neel
temperature. Inset: inverse susceptibility and the high-T
Curie-Weiss fit (solid line).}
\end{figure}

\begin{figure}
\caption{\label{fig.TOMU03}Low temperature susceptibility of
ScMnO$_3$ in a field of 2~T. The kink represents the Neel
temperature. Inset: inverse susceptibility and the high-T
Curie-Weiss fit (solid line).}
\end{figure}

\begin{figure}
\caption{ \label{fig.TOMU04}Excess specific heat (left scale) of
YMnO$_3$ after subtraction of phonon contribution (see inset).
There is no change in a field of 9~T. The solid line represents
the entropy (right scale). T$_N$ is given by the peak in C$_p$(T).
The dashed line in the inset shows the lattice contribution. }
\end{figure}

\begin{figure}
\caption{\label{fig.TOMU05}Excess specific heat (left scale) of
LuMnO$_3$ after subtraction of phonon contribution (see inset).
There is no change in a field of 9~T. The solid line represents
the entropy (right scale). T$_N$ is given by the peak in C$_p$(T).
The dashed line in the inset shows the lattice contribution.}
\end{figure}

\begin{figure}
\caption{ \label{fig.TOMU06}Excess specific heat (left scale) of
ScMnO$_3$ after subtraction of phonon contribution (see inset).
There is no change in a field of 9~T. The solid line represents
the entropy (right scale). T$_N$ is given by the peak in C$_p$(T).
The dashed line in the inset shows the lattice contribution.}
\end{figure}

\begin{figure}
\caption{ \label{fig.TOMU07}Dielectric constant $\epsilon$ around
the magnetic phase transition. The arrows represents T$_N$
determined from the $\chi$ and $\Delta$C$_p$ measurements. }
\end{figure}

\begin{figure}[t]
\caption{\label{fig.TOMU08}ZFC-FC susceptibility of ScMnO$_3$ in
0.05~T and 2~T. Inset: low temperature specific heat of YMnO$_3$,
LuMnO$_3$ and ScMnO$_3$ on logarithmic scales.}
\end{figure}


\begin{references}
\bibitem{Yakel} Yakel H L, Koehler W C, Bertaut E F and Forrat E F 1963 {\it Acta Cryst.} {\bf 16} p~957--962

\bibitem{koehler} Koehler W C, Yakel H L, Wollen E O and Cable J W 1964 {\it Phys. Lett.} {\bf 9} 93

\bibitem{WA} Waintal A and Chenevas J 1967 {\it Mat. Res. Bull.} {\bf 2}
p~819--822

\bibitem{McCarty} McCarty G J, Gallagher P V and Sipe C 1973 {\it Mat. Res. Bull.} {\bf 8}
p~1277--84

\bibitem{smolenski} Smolenskii G A and Chupis I E 1982 {\it Sov. Phys. Usp.} {\bf 25} p~475--493

\bibitem{fujimura} Fujimura N, Ishida T, Yoshimura T and Ito T 1996 {\it Appl. Phys. Lett.} {\bf 69}
p~1011--13
\bibitem{Bieringer99} Bieringer M and Greedan J E 1999 {\it J. Solid State Chem.} {\bf 143} p~132--139

\bibitem{Munoz00} Mu\~{n}oz A, Alonso J A, Mart\'{i}nez-Lope M J, Cas\'{a}is M T, Mart\'{i}nez J L and Fern\'{a}ndez-D\'{i}az  2000
{\it Phys. Rev. B.} {\bf 62} p~9498--9510

\bibitem{Hill00} Hill N 2000 {\it J. Phys. Chem. B} {\bf 104}
p~6694--6709

\bibitem{wan} Wan X, Dong J, Qian M and Zhang W 2000 {\it Phys. Rev. B} {\bf 61} p~10664--669

\bibitem{Xu} Xu H W, Iwasaki J, Shimizu T, Satoh H and Kamegashira N 1995 {\it J. Alloys and Compounds} {\bf 221} p~274--279

\bibitem{Huang} Huang Z J, Cao Y, Sun Y Y, Xue Y Y and Chu C W 1997 {\it Phys. Rev. B.} {\bf 56} p~2623--26

\bibitem{AA}Alfa Aesar Johnson Matthey GmbH, Postfach 11 07 65, D-76057 Karlsruhe, Germany, dcat@alpfa.com

\bibitem{karen98} Karen P and Woodward P M 1998 {\it J. Solid State Chem.} {\bf 141} p~78--88

\bibitem{alpha} Bertaut E F and Mercier M 1963 {\it Phys. Lett.} {\bf 5}
p~27--29

\bibitem{ramirez}Ramirez A P 2001 {\it Geometrical Frustration}, to be published in
{\it Handbook of Magnetic Materials} ed. K.H.J. Buschow
(Amsterdam: North-Holland/Elsevier B C)


\bibitem{iliev} Iliev M N, Lee H -G, Popov V N, Abrashev M V, Hamed A, Meng R L and Chu C W 1997 {\it Phys. Rev. B.} {\bf 56} p~2488--94

\bibitem{frolich} Fr\"{o}hlich D, Leute St, Pavlov V V and Pisarev R V 1998 {\it Phys. Rev. Lett.} {\bf 81} p~3239--42

\bibitem{frolichPRL} Fiebig M, Fröhlich D, Kohn K, Leute St, Lottermoser Th, Pavlov V V and Pisarev R V 2000 {\it Phys. Rev. Lett.} {\bf 84} p~5620-5623


\end{references}
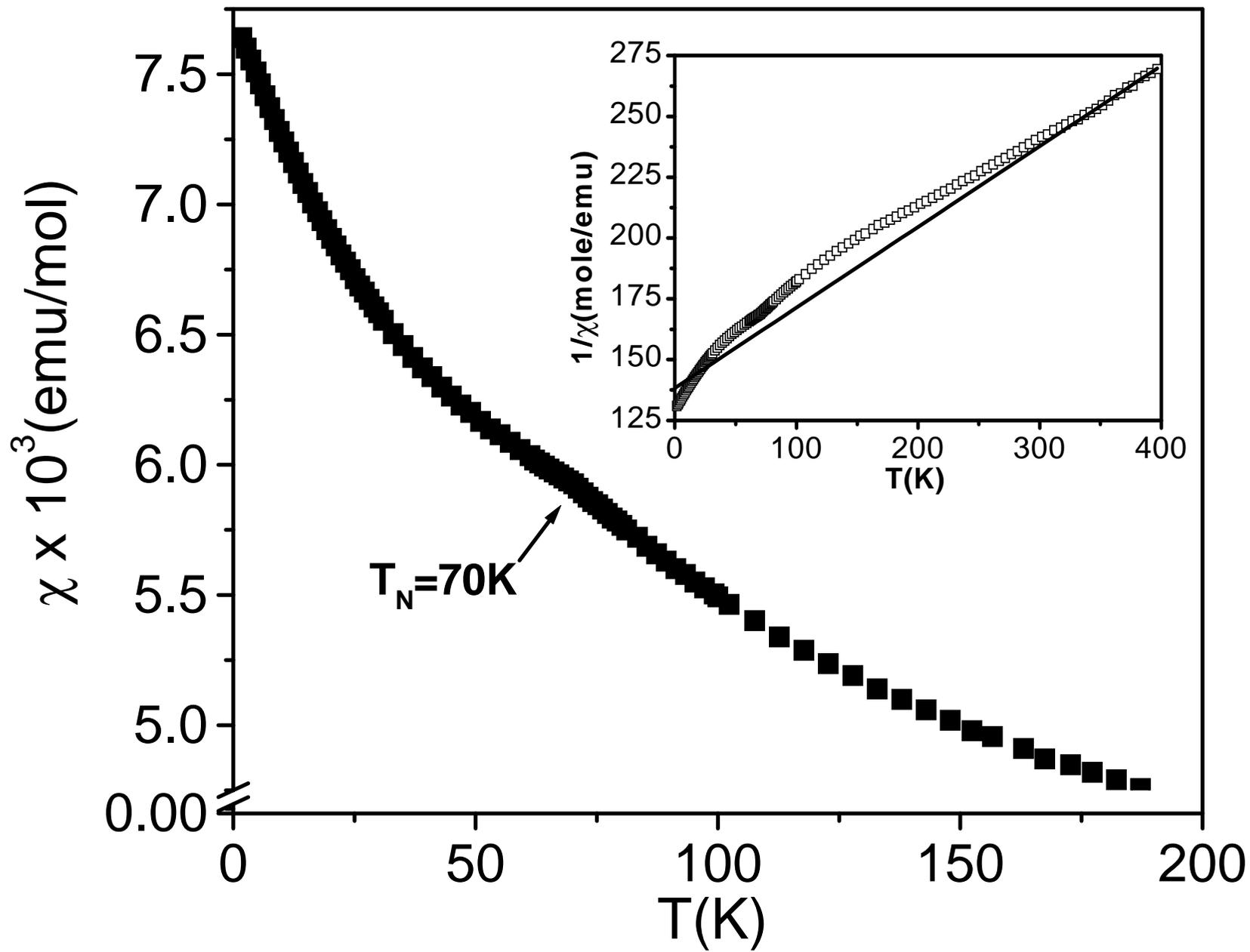
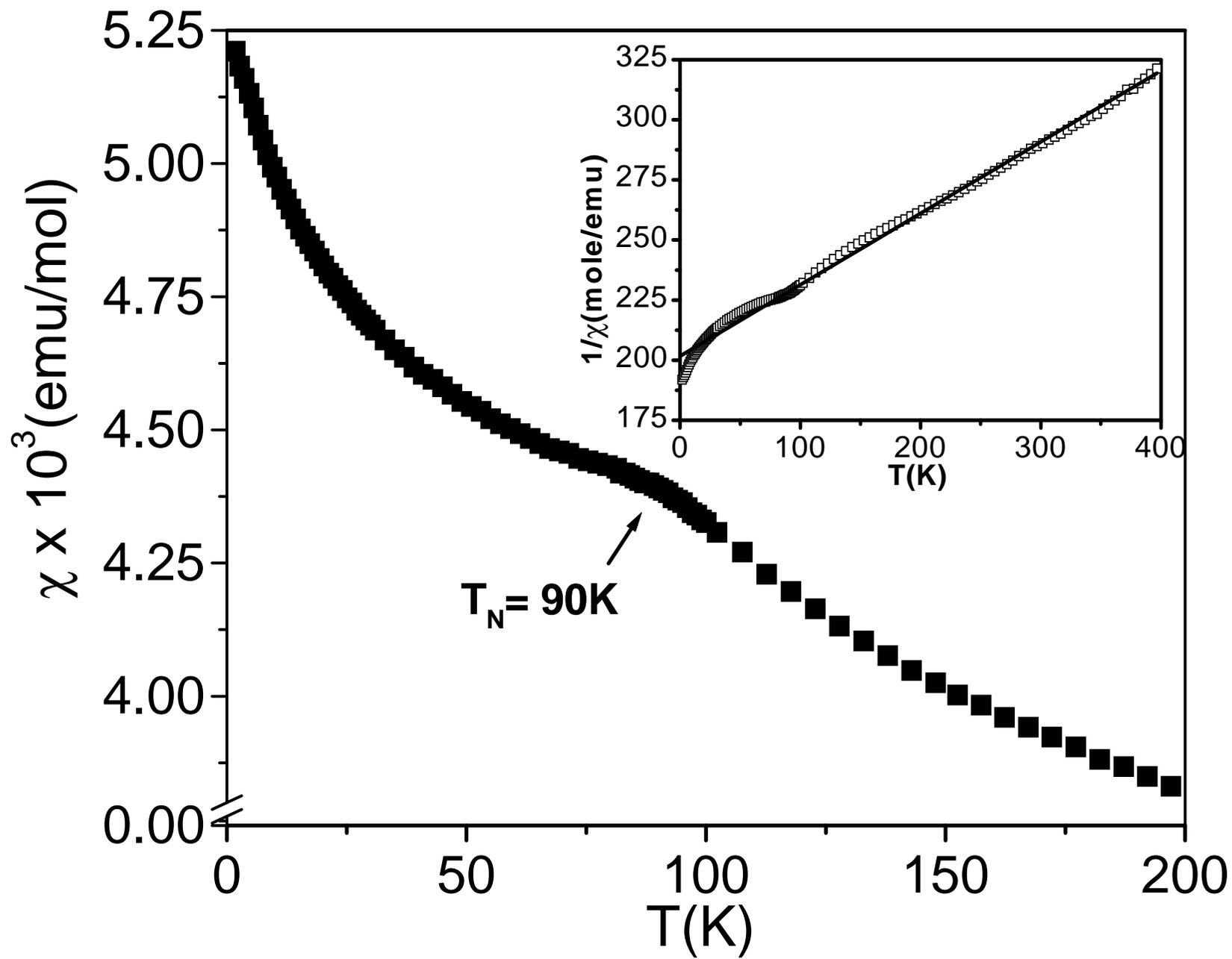
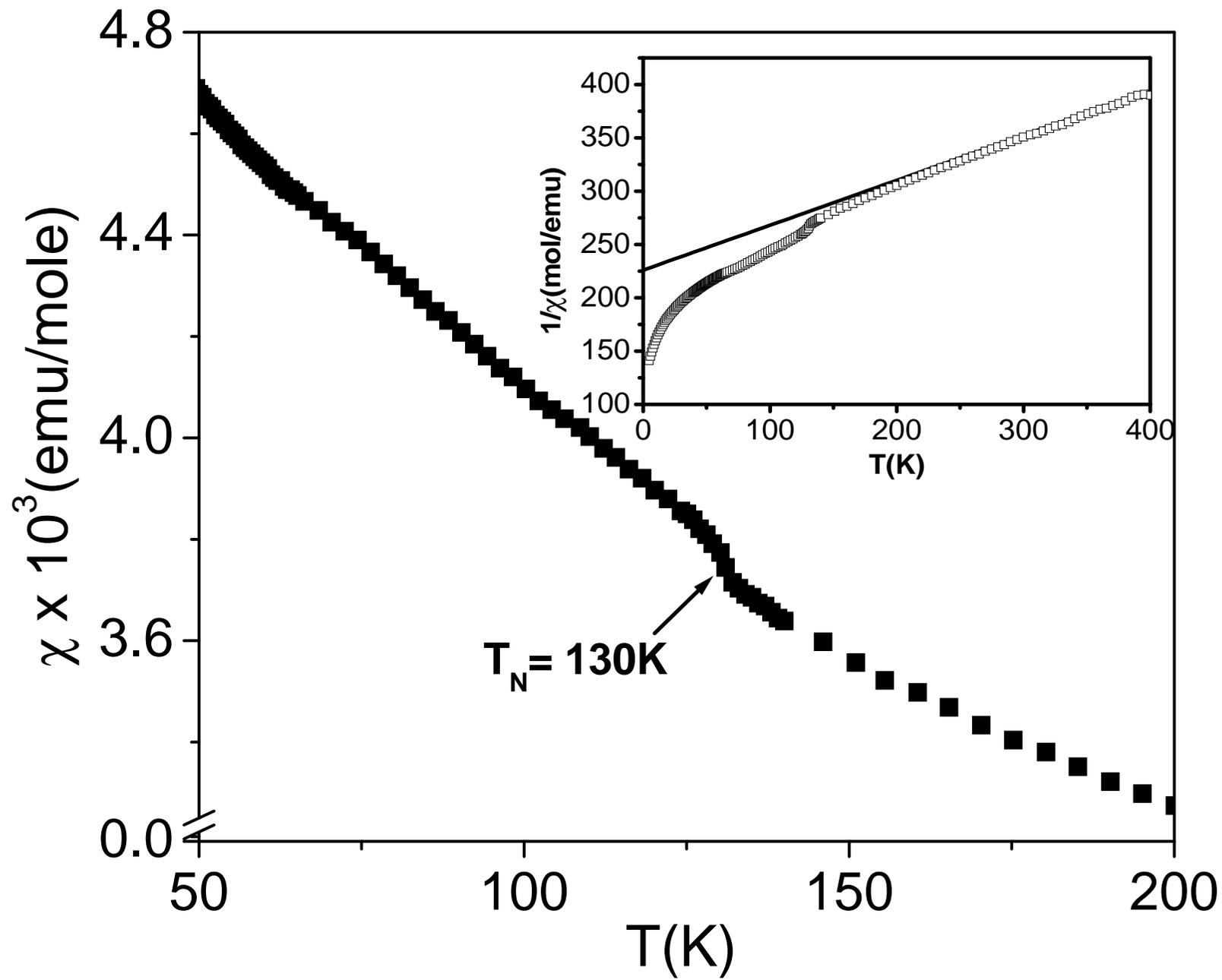
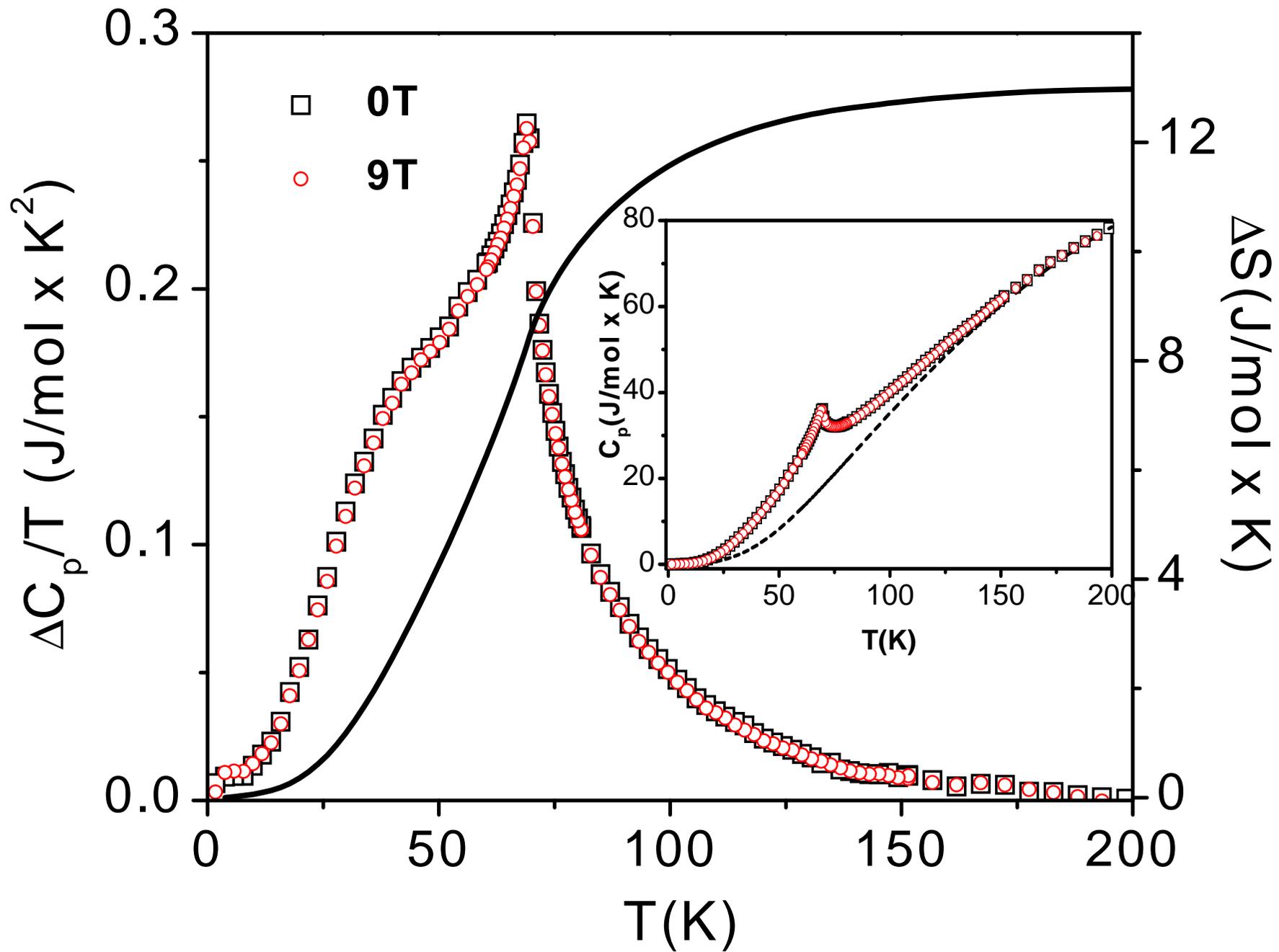
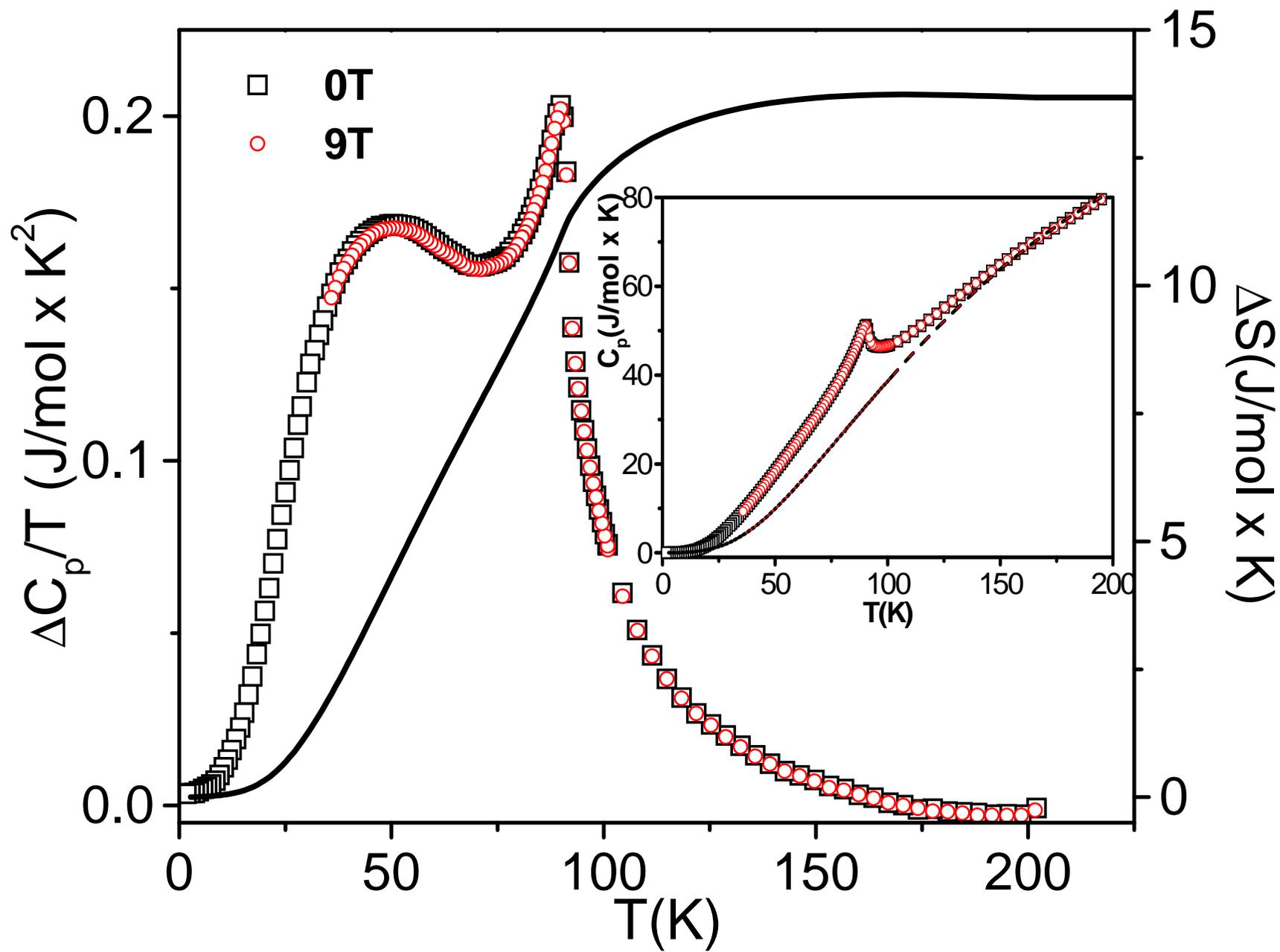
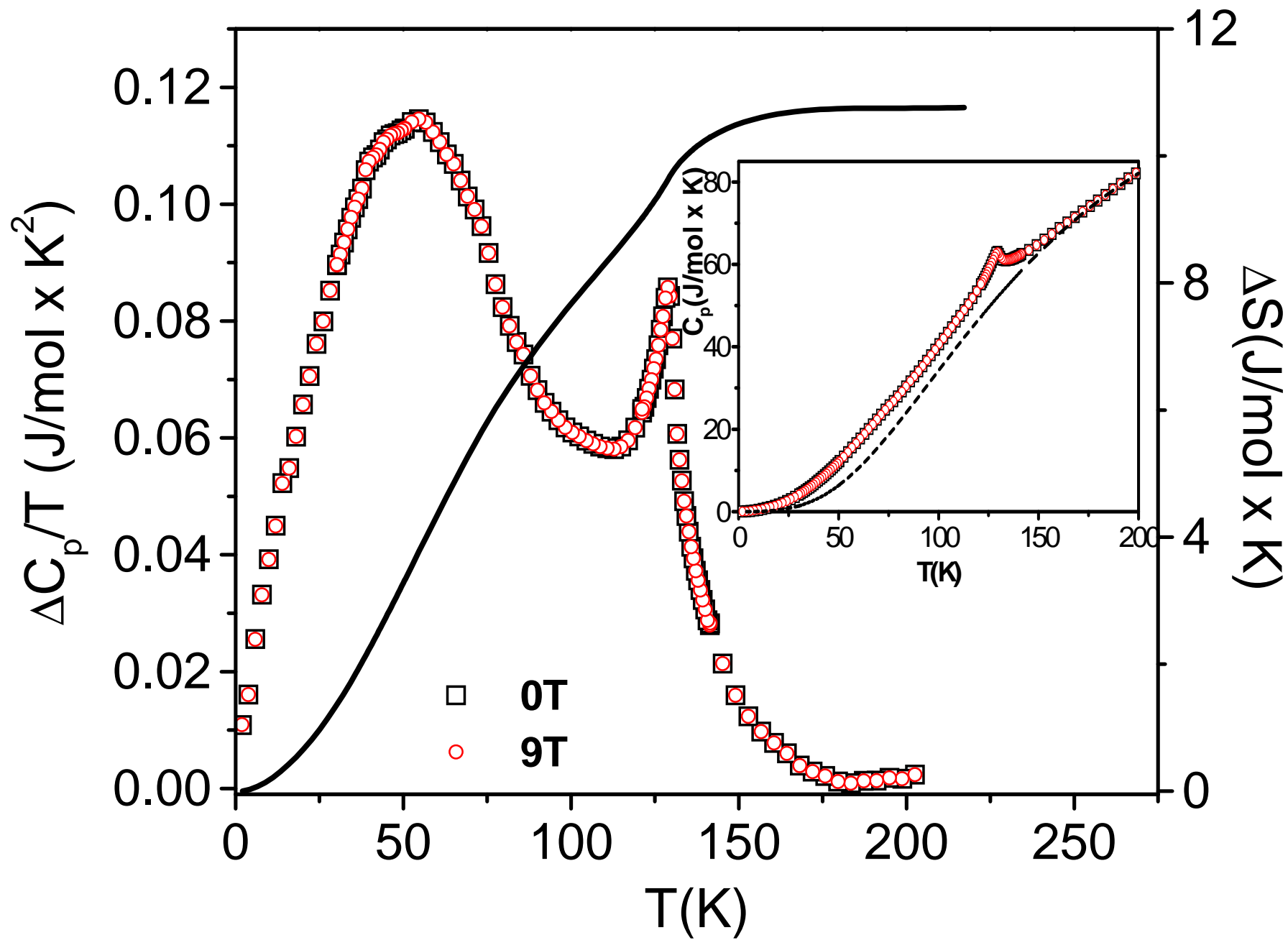
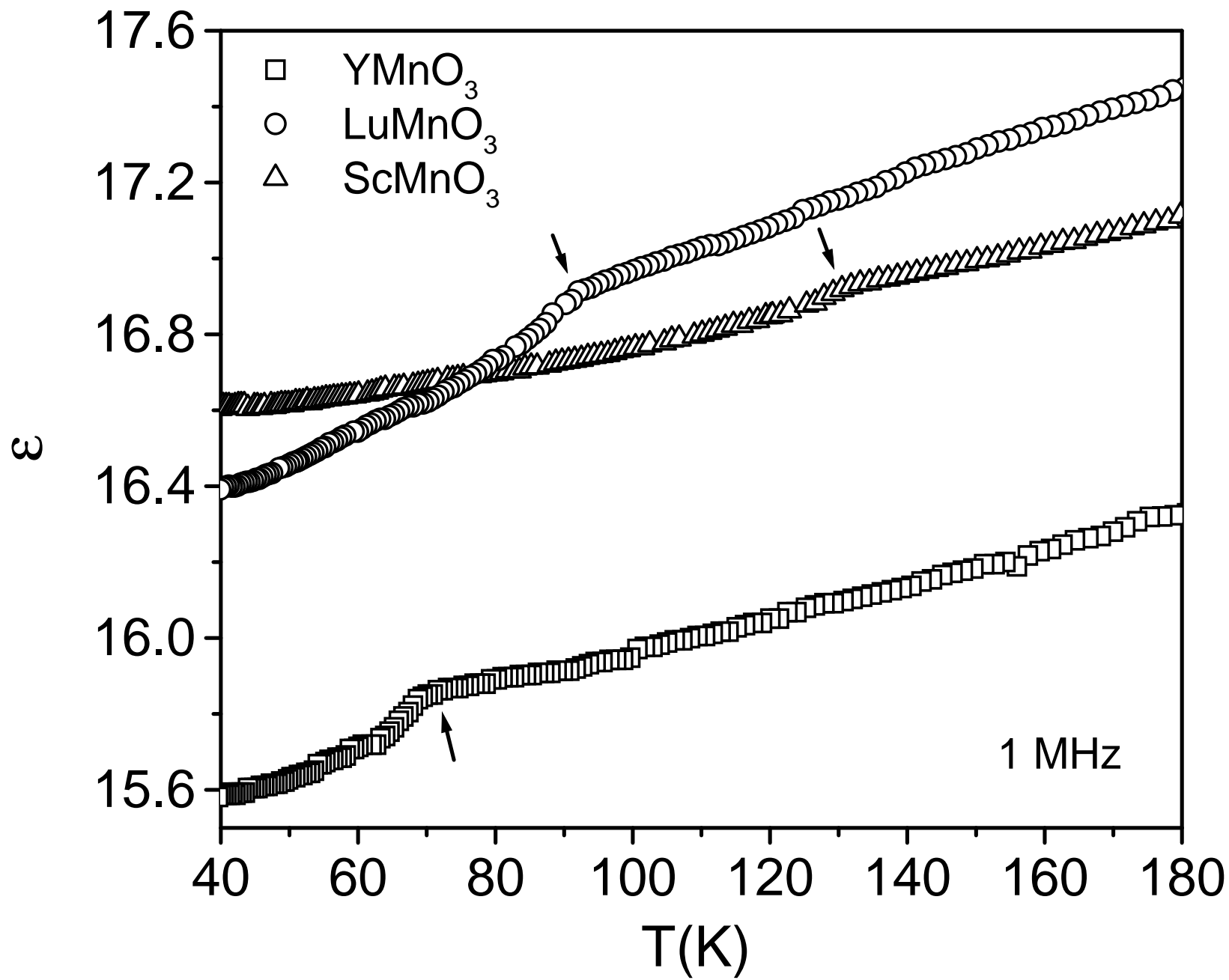
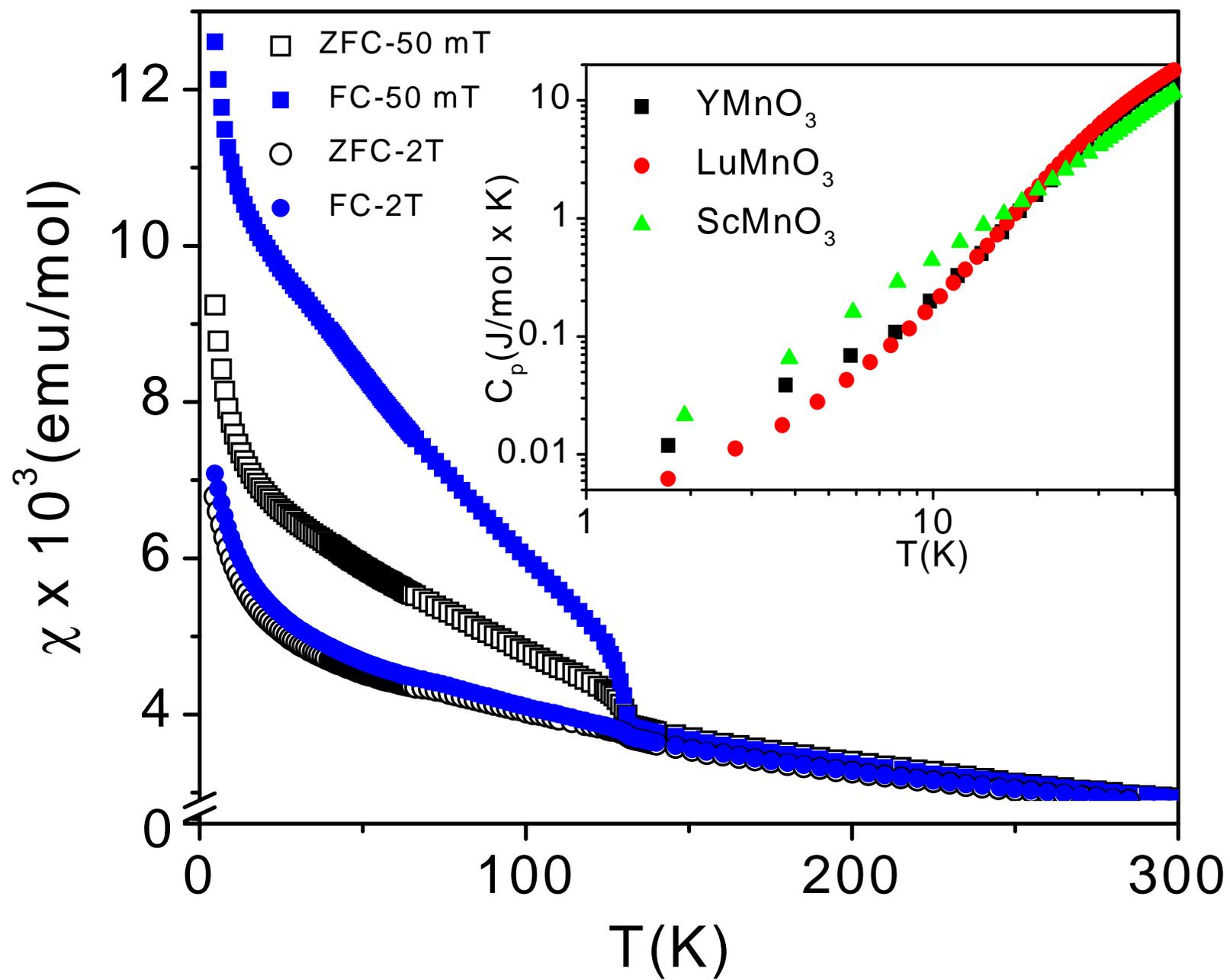
\end{document}